\documentclass[aps,pra,twocolumn,showpacs,10pt]{revtex4-1}

\usepackage{epsfig,amssymb,amsmath,graphicx,psfrag,color}
\usepackage[dvips]{hyperref}

\begin{document}

\title{Light scattering from ultracold atomic gases in optical lattices at finite temperature}
\author{James S. Douglas$^1$ and Keith Burnett$^2$}
\affiliation{$^1$Clarendon Laboratory, University of Oxford, Oxford OX1 3PU, United Kingdom
\\$^2$University of Sheffield, Western Bank, Sheffield S10 2TN, United Kingdom}
\begin{abstract}
We study light scattering from atoms in optical lattices at finite temperature. We examine the light scattered by fermions in the noninteracting regime and by bosons in the superfluid and Mott insulating regimes. We extend previous theoretical studies to include the full band structure of the optical lattice. We find that light scattering that excites atoms out of the lowest band leads to an increase in light scattering away from the classical diffraction peaks and is largely temperature independent. This additional light scattering leads to lower efficiency of temperature measurements based on photon counting.
\end{abstract}
\pacs{37.10.Jk  42.50.Ct  37.10.Vz}
\maketitle

Ultracold atoms in optical lattices exhibit a remarkable range of behavior and have become an indispensable tool for investigating many-body phenomena \cite{Lewenstein2007a,Bloch2008a}. As the gamut of many-body states achieved in experiments with optical lattices expands, it is also important to consider how to detect and reveal more information about these states. With this goal in mind, many authors have considered the scattering of light by these ultracold gases as a potential way of extracting information.

Bragg scattering, where light scattering from the sample is stimulated by a pair of lasers, is a prominent method that has been investigated as a probe of bosons in optical lattices \cite{Menotti2003a,Oosten2005a,Rey2005a,Rist2010a,Ye2011a}, and experiments using Bragg scattering have detected signatures of the superfluid and Mott insulator states of the Bose-Hubbard model \cite{Clement2009a,Ernst2010a}.
Another suggested scheme involves light scattering into a cavity, where the strong light-matter coupling encodes information about the many-body state into the distribution of photons transmitted by the cavity  \cite{Mekhov2007a,Mekhov2007b,Mekhov2007c,Chen2007a,Chen2009a}. The effect of finite temperature on light scattering has been investigated as a potential probe of temperature for the superfluid and Mott cases \cite{Trippenbach2009a} and also for the noninteracting Fermi-Dirac gas \cite{Ruostekoski2009a}. Recently the far field diffraction pattern of a Mott insulator was imaged in an experiment where the diffraction peaks were used to detect the spin distribution \cite{Weitenberg2011a}. Detection of spin distributions in optical lattices using light-matter interactions has also been considered theoretically \cite{Cherng2007a,Eckert2007a,Vega2008a,Eckert2008a,Roscilde2009a,Javanainen2003,Ruostekoski2008a,Douglas2010a,Corcovilos2010a,Chiara2011a}.

\begin{figure}
\centering
\includegraphics{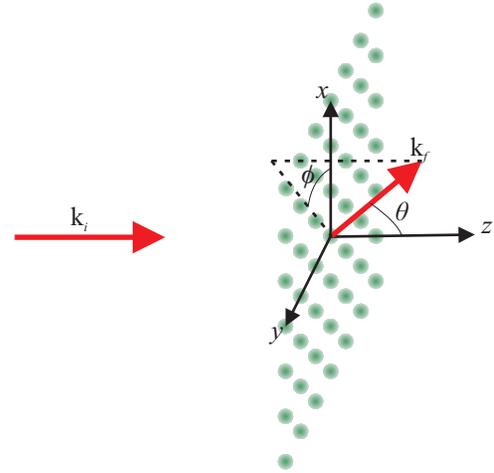}
\caption{(Color online) Light scattering scheme. The atoms in the optical lattice are illuminated by a laser beam with wave vector $\mathbf{k}_i$. Light is then scattered to other modes $\mathbf{k}_f$ with spherical coordinates $\theta$ and $\phi$. In our examples of light scattering the optical lattices are two-dimensional square lattices with lattice translation vectors $a\mathbf{\hat{x}}$ and $a\mathbf{\hat{y}}$, and the input light has $\mathbf{k}_i = \pi\mathbf{\hat{z}}/a$. In this coordinate system we then have $\mathbf{k} = \mathbf{k}_i-\mathbf{k}_f = -\pi(\sin\theta\cos\phi,\sin\theta\sin\phi,\cos\theta-1)/a$.}\label{fig:light_scatter}
\end{figure}

In this paper we consider what light scattering, which is not stimulated by another laser or by a cavity, tells us about the many-body state. The aim being to reveal information about the atoms while they remain \textit{in situ} with a relatively simple experimental implementation, only requiring a laser to be shone on the atoms, as depicted in Fig.~\ref{fig:light_scatter}, followed by the detection of scattered photons, possible after collection by a lens. This would remove the need for a cavity or the need to destructively image the atomic sample, which is usually required with Bragg scattering (see Ref.~\cite{Pino2011a} for a recent exception). When scattering is not stimulated there is less restriction on the final energy state of the scattered photons and the full band structure of the optical lattice must be taken into account, something which has been neglected in previous works. We present here a treatment of light scattering from ultracold atoms in optical lattices that includes both the band structure and the effects of nonzero temperature.

Our approach is organized as follows. We begin by describing the light scattering process that forms the basis of our work. We then examine the light scattering patterns that result for noninteracting fermions, and for bosons in the superfluid and Mott insulator regimes of the Bose-Hubbard model. We then examine how the temperature measurement described by Ruostekoski \textit{et al.}~\cite{Ruostekoski2009a} applies to these systems and is affected by the multiband structure of the optical lattice.


\section{First order light scattering}

Nonresonant light scattering from an ultracold gas can occur in two ways. The first involves the diffraction of light due to the nonuniform density of the gas, which leaves the gas unchanged, while the second involves the creation of excitations in the gas where the light transfers momentum and/or energy to the gas. In the following we see how these processes occur when laser light is scattered from atoms in first-order perturbation theory.

We consider light-matter interactions where the atom can be treated as a two-level system, as may be the case because of angular momentum considerations in the dipole approximation.
For example, in $^{87}$Rb, the transition $|5^2S_{1/2},F=2,m_F=2\rangle \rightarrow |5^2P_{3/2}, F=3, m_F=3\rangle$ that is excited by $\sigma^+$-polarized light is a cycling transition, where, because of the polarization of the input light, only one excited state can take part, which in turn can only decay to the initial ground state. 
Further simplification can be made when the light incident on the atoms is off-resonant, with detuning $\Delta = \omega_a-\omega_i$ between the atomic resonant frequency $\omega_a$ and the input laser frequency $\omega_i$. In which case the excited atomic state can be adiabatically eliminated, giving, in the rotating wave approximation, the following interaction Hamiltonian,
\begin{equation}
H= -g\int d\mathbf{r}\hat{\Psi}^\dagger(\mathbf{r})
\mathbf{\tilde{E}}^-(\mathbf{r})\cdot\mathbf{\tilde{E}}^+(\mathbf{r})\hat{\Psi}(\mathbf{r}),
\end{equation}
where $g = 3\pi\epsilon_0c^3 \gamma/(\omega_a^3\Delta)$ and $\gamma$ is the lifetime of the excited state. The Hamiltonian is expressed in terms of the ground-state atomic-field operators $\hat{\Psi}^\dagger(\mathbf{r})$ and $\hat{\Psi}(\mathbf{r})$ and the positive and negative frequency components of the electric field, $\mathbf{\tilde{E}}^+(\mathbf{r})$ and $\mathbf{\tilde{E}}^-(\mathbf{r})$.

In the interaction picture we have $\hat{H}_I(t) = e^{i \hat{H}_0 t/\hbar}\hat{H}e^{-i \hat{H}_0 t/\hbar}$, where $\hat{H}_0$ is the Hamiltonian of the noninteracting light-matter system, and the time evolution operator is given by the integral equation $U_I(t,0) = 1 -\frac{i}{\hbar}\int^t_0 dt'\hat{H}_{I}(t')U_I(t',0)$ \cite{FW71}. For a weak perturbation the evolution of the state is given approximately by expanding this integral equation to first order, giving
\begin{equation}
|\Psi(t)\rangle_I=\left(1-\frac{i}{\hbar}\int^t_0 dt'\hat{H}_{I}\right)|\Psi(0)\rangle_I.
\end{equation}

We take the initial state of the system to be an atomic eigenstate $|\Psi_u\rangle$ and a laser represented by the  classical field $\mathbf{\tilde{E}}^+(\mathbf{r}) = \frac{\mathcal{E}}{2}e^{i\mathbf{k}_i\cdot\mathbf{r}}\boldsymbol{\epsilon}_{i}$. The first-order coupling then leads to photons being scattered from the laser into other modes $(\mathbf{k}_f,\lambda)$ with wave vector $\mathbf{k}_f$, frequency $\omega_f$ and polarization $\boldsymbol{\epsilon}_\lambda(\mathbf{k}_f)$, resulting in a momentum change $\hbar\mathbf{k}=\hbar(\mathbf{k}_i-\mathbf{k}_f)$ and energy change $\hbar\omega=\hbar(\omega_{i}-\omega_{f})$. The probability at time $t$ of the system being in a new atomic eigenstate $|\Psi_v\rangle$ with an additional photon in the mode $(\mathbf{k}_f,\lambda)$
is then
\begin{multline}
\left|\left\langle \mathbf{k}_f, \lambda \left|\otimes \left\langle\Psi_v\left| \frac{1}{\hbar}\int^t_0 dt'\hat{H}_{I}\right|\Psi_u\right\rangle\otimes \right|0\right\rangle\right|^2 
= 
 t G_\lambda(\mathbf{k}_f) \\\times
\delta_t((E_v-E_u)/\hbar-\omega)
\left|\int d\mathbf{r}e^{i\mathbf{k}\cdot \mathbf{r}}\langle\Psi_v| \hat{\Psi}^\dagger(\mathbf{r})\hat{\Psi}(\mathbf{r})|\Psi_u\rangle\right|^2
\end{multline}
where $G_\lambda(\mathbf{k}_f)= |g\mathcal{E}\boldsymbol{\epsilon}_\lambda^*(\mathbf{k}_f)\cdot\boldsymbol{\epsilon}_i|^2 \omega_f/(8\hbar\epsilon_0(2\pi)^2)$ is the coupling constant between the electromagnetic field modes  and $E_{j}$ is the unperturbed energy of the state $|\Psi_j\rangle$. The function $\delta_t(\omega) =  2 \sin^2(\omega t/2)/(\pi t\omega^2)$ approaches the Dirac delta function $\delta(\omega)$ as the interaction time $t$ approaches infinity, enforcing energy conservation.

For $\sigma^+$-polarized input laser light we can sum over polarizations in the coupling to get
$G(\mathbf{k}_f) = \sum_\lambda G_\lambda(\mathbf{k}_f)= |g\mathcal{E}|^2(1+\cos(\theta)^2) \omega_f/(16\hbar\epsilon_0(2\pi)^2)$.
The total rate of scattering photons with wave vector $\mathbf{k}_f$ and frequency $\omega_f$ of either polarization is then
\begin{equation}
\Gamma(\mathbf{k}_f,\omega_f)=G(\mathbf{k}_f)S(\mathbf{k}_i-\mathbf{k}_f,\omega_i-\omega_f),
\label{eq:scattering_rate}
\end{equation}
where $S(\mathbf{k},\omega)$ is the dynamic structure factor \cite{VanHove1954}.
For a finite-temperature system, in the canonical ensemble with partition function $Z$, the dynamic structure factor is \cite{VanHove1954,Griffin1993a}
\begin{multline}
S(\mathbf{k},\omega)=\frac{1}{Z}\sum_{u,v} \delta((E_v-E_u)/\hbar-\omega)e^{-E_u/(k_B T)}
\\\times \left|\int d\mathbf{r}e^{i\mathbf{k}\cdot \mathbf{r}}\langle\Psi_v| \hat{\Psi}^\dagger(\mathbf{r})\hat{\Psi}(\mathbf{r})|\Psi_u\rangle\right|^2 .
\end{multline}
The structure factor can be divided up into two parts, the first with $u=v$ describes classical diffraction, which results in no energy or momentum transfer to the gas, that is, the light scattering is \textit{elastic}. The second part, where $u\neq v$, describes \textit{inelastic} light scattering that results in excitations of the gas. 

For nonresonant light scattering the frequency change $\omega$ is determined by the difference in energies between the initial and final atom many-body states. For scattering from ultracold atoms these energy differences are all many orders of magnitude less than the frequency of the input light. To a very good approximation we then have $\mathbf{k}_f = |\mathbf{k}_i|\hat{\mathbf{k}}_f$, where $\hat{\mathbf{k}}_f$ has associated angles $\theta$ and $\phi$ as shown in Fig.~\ref{fig:light_scatter}. 
To get the total rate of photon scattering in direction $\hat{\mathbf{k}}_f$, we integrate Eq.~(\ref{eq:scattering_rate}) over frequency. This gives
\begin{equation}
I_{total}(\hat{\mathbf{k}}_f)=I_{atom}(\theta)\times S(\mathbf{k}_i-|\mathbf{k}_i|\hat{\mathbf{k}}_f),
\end{equation}
where
$I_{atom}(\theta) = 9 I_{in}  \gamma^2(1+\cos(\theta)^2)/(32\hbar c k_i^3 \Delta^2)$
is the scattering distribution resulting from the electronic structure of the atom and laser intensity $I_{in} = \epsilon_0 c|\mathcal{E}|^2/2$, and
\begin{multline}
S(\mathbf{k}) = \frac{1}{Z}\sum_u\int d\mathbf{r}d\mathbf{r'}e^{i\mathbf{k}\cdot (\mathbf{r}-\mathbf{r'})} e^{-E_u/(k_B T)}\\
\times\langle\Psi_u| \hat{\Psi}^\dagger(\mathbf{r'})\hat{\Psi}(\mathbf{r'})\hat{\Psi}^\dagger(\mathbf{r})\hat{\Psi}(\mathbf{r})|\Psi_u\rangle
\end{multline}
is the scattering distribution due to the spatial structure of the atomic sample, known as the static structure factor \cite{Griffin1993a}. 

The static structure factor applies in the perturbative regime, where it has been used successfully to describe neutron scattering from liquid helium \cite{VanHove1954} and to calculate the spectrum of light scattered from ultracold gases \cite{Javanainen1995b,Graham1996a}, along with the response of ultracold gases to Bragg excitation \cite{Ozeri2005a}. In this paper we use the structure factor to predict the angular dependence of light scattering for atoms in an optical lattice. While our formalism can be applied to any lattice and input laser configuration, in all our examples we consider the setup in Fig.~\ref{fig:light_scatter}, where the optical lattice is assumed to be a two-dimensional square lattice in the $xy$ plane, illuminated by light propagating in the $z$ direction with wavelength equal to that of the light used to create the optical lattice. The change in photon wavevector upon scattering is then $\mathbf{k} = \mathbf{k}_i-\mathbf{k}_f = -\pi(\sin\theta\cos\phi,\sin\theta\sin\phi,\cos\theta-1)/a$, where $\phi$ and $\theta$ are the scattering angles shown in Fig.~\ref{fig:light_scatter} and $a$ is the lattice intersite separation. We denote the strength of the optical lattice potential in the $x$, $y$ and $z$ directions by $V_x$, $V_y$, and $V_z$, respectively, in units of the recoil energy $E_R = \frac{\hbar^2\pi^2}{2ma^2}$.
In the following sections we calculate the light scattering predicted for fermions and bosons and see how different correlations in different many-body states affect the angular distribution of the scattered light.


\section{Fermions}

For spin-polarized fermions of the same species, $s$-wave scattering is prohibited, and for low temperatures the fermionic atoms are approximately noninteracting. The Hamiltonian for the many-body system is then exactly diagonalized by the single-particle Bloch states $\phi_\mathbf{q,m}(\mathbf{r})$\cite{Kittel}, where $\mathbf{q}$ is the quasimomentum and $\mathbf{m}$ is the band index. At finite temperature each Bloch state is occupied according to Fermi-Dirac statistics.

We can expand the atomic-field operators in terms of the Bloch states, $\hat{\Psi}(\mathbf{r}) = \sum_{\mathbf{q},\mathbf{m}}\hat{b}_{\mathbf{q},\mathbf{m}}\phi_{\mathbf{q},\mathbf{m}}(\mathbf{r})$, where $\hat{b}_{\mathbf{q},\mathbf{m}}$ is the annihilation operator of an atom in the Bloch state $\phi_{\mathbf{q},\mathbf{m}}(\mathbf{r})$. Together with the corresponding set of creation operators $\hat{b}^\dagger_{\mathbf{q},\mathbf{m}}$, these operators obey the standard fermion anticommutation relations.
This expansion leads to the following expression for the static structure factor,
\begin{multline}
S( \mathbf{k}) = \sum_{\substack{\mathbf{q_1},\mathbf{q_2},\mathbf{q_3},\mathbf{q_4}\\\mathbf{m_1},\mathbf{m_2},\mathbf{m_3},\mathbf{m_4}}}\!\!f^*_{\mathbf{q}_1,\mathbf{m_1},\mathbf{q}_2,\mathbf{m_2}}( \mathbf{k})f_{\mathbf{q}_4,\mathbf{m_4},\mathbf{q}_3,\mathbf{m_3}}(\mathbf{k})\\\times 
\langle\hat{b}_{\mathbf{q_1},\mathbf{m_1}}^\dagger\hat{b}_{\mathbf{q_2},\mathbf{m_2}}
\hat{b}^\dagger_{\mathbf{q_3},\mathbf{m_3}}\hat{b}_{\mathbf{q_4},\mathbf{m_4}}\rangle,
\label{eq:Static_fact_expand}
\end{multline}
where
\begin{equation}
f_{\mathbf{q}_1,\mathbf{m}_1,\mathbf{q}_2,\mathbf{m}_2}(\mathbf{k}) = M\int d\mathbf{r} \phi_\mathbf{q_2,m_2}^*(\mathbf{r})\phi_\mathbf{q_1,m_1}(\mathbf{r})
e^{i \mathbf{k}\cdot\mathbf{r}}
\end{equation}
is the transition matrix element between the Bloch functions associated with a momentum transfer $\mathbf{k}$.
For sufficiently low temperatures all the atoms reside initially in the lowest band of the optical lattice; however atoms may be excited to higher bands by scattering a photon. 
We then have two terms, the first due to light scattering within the lowest band of the optical lattice, that is,
\begin{multline}
S_g (\mathbf{k})= \sum_{\mathbf{q_1},\mathbf{q_2},\mathbf{q_3},\mathbf{q_4}}\!\!f^*_{\mathbf{q}_1,\mathbf{0},\mathbf{q}_2,\mathbf{0}}( \mathbf{k})f_{\mathbf{q}_4,\mathbf{0},\mathbf{q}_3,\mathbf{0}}( \mathbf{k})\\\times
\langle\hat{b}_{\mathbf{q_1},\mathbf{0}}^\dagger\hat{b}_{\mathbf{q_2},\mathbf{0}}
\hat{b}^\dagger_{\mathbf{q_3},\mathbf{0}}\hat{b}_{\mathbf{q_4},\mathbf{0}}\rangle ,
\end{multline}
and the second due to scattering into the higher bands of the lattice, that is
\begin{equation}
S_b(\mathbf{k}) =  \!\!\sum_{\mathbf{q_1},\mathbf{q_2},\mathbf{q_3},\mathbf{m}\neq\mathbf{0}}\!\!
 f^*_{\mathbf{q}_1,\mathbf{0},\mathbf{q}_2,\mathbf{m}}( \mathbf{k})f_{\mathbf{q}_3,\mathbf{0},\mathbf{q}_2,\mathbf{m}}( \mathbf{k})\langle\hat{b}_{\mathbf{q_1},\mathbf{0}}^\dagger\hat{b}_{\mathbf{q_3},\mathbf{0}}\rangle.
\end{equation}
This part of the structure factor was neglected in other treatments of finite-temperature light scattering \cite{Trippenbach2009a,Ruostekoski2009a}, and we find that this can make a significant difference to predictions about the light scattering.

We can calculate the higher band component with the help of the sum rule
\begin{equation}
\sum_{\mathbf{q},\mathbf{m}}f^*_{\mathbf{q}_1,\mathbf{0},\mathbf{q},\mathbf{m}}( \mathbf{k})f_{\mathbf{q}_2,\mathbf{0},\mathbf{q},\mathbf{m}}( \mathbf{k}) = \delta_{\mathbf{q_1},\mathbf{q_2}},
\end{equation}
which is implied by the completeness of the Bloch functions. Using this relation we find
\begin{equation}
S_b(\mathbf{k}) =  N - \sum_{\mathbf{q_1},\mathbf{q_2},\mathbf{q_3}}\!\!
  f^*_{\mathbf{q}_1,\mathbf{0},\mathbf{q}_2,\mathbf{0}}( \mathbf{k})f_{\mathbf{q}_3,\mathbf{0},\mathbf{q}_2,\mathbf{0}}( \mathbf{k})\langle\hat{b}_{\mathbf{q_1},\mathbf{0}}^\dagger\hat{b}_{\mathbf{q_3},\mathbf{0}}\rangle,
\end{equation}
where $N$ is the total number of atoms on our $M$-site lattice. All operators are now in the lowest band and we drop the band index from our notation.


\begin{figure}
\centering
\includegraphics{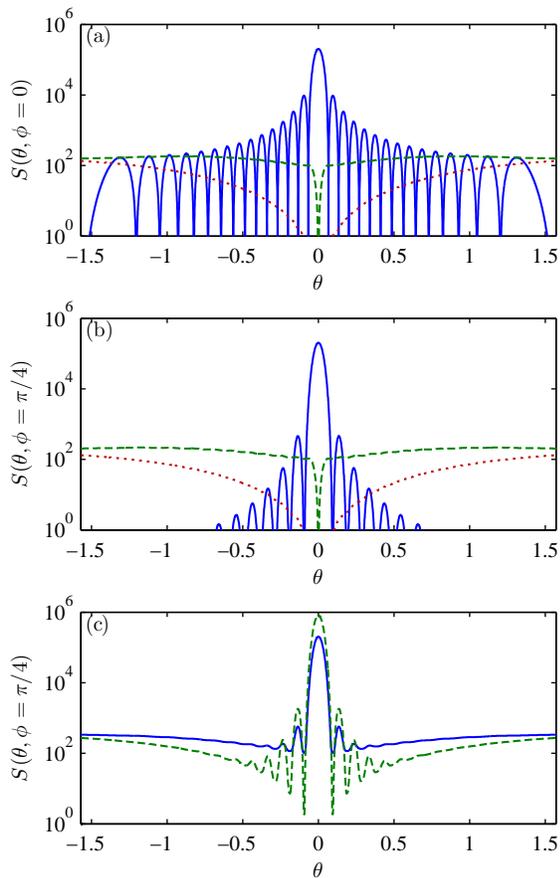}
\caption{(Color online) Comparison of the angular dependence of the components of the static structure factor for a Fermi-Dirac gas on a $30\times30$ lattice with $V_x=V_y=8 E_R$, $V_z = 15 E_R$ and temperature $0.02 E_R/k_B = 0.16 T_F$, where $T_F$ is the Fermi temperature.
(a) Dependence on $\theta$ of the structure factor components $S_{g0}(\mathbf{k})$ (solid blue line), $S_{g1}(\mathbf{k})$ (dashed green line)  and $S_b(\mathbf{k})$ (dotted red line) for $\phi=0$ and filling factor $f = 0.5$. (b) As in panel (a) with $\phi=\pi/4$. (c) Comparison of the structure factors for gases with filling factors $f = 0.5$ (solid blue line) and $f=1$ (dashed green line).
}\label{fig:Fermion_sf}
\end{figure}


We can now evaluate all the expectation values using Fermi-Dirac statistics. We find that light scattering within the lowest band is the sum of two components $S_g (\mathbf{k})=S_{g0} (\mathbf{k})+S_{g1} ( \mathbf{k})$. The zeroth-order component, which does not result in many-body excitations in the lattice, is
\begin{multline}
S_{g0} (\mathbf{k})=\left|\sum_{\mathbf{q}}f_\mathbf{q,q}(\mathbf{k})
N_\mathbf{q} \right|^2 
+\sum_{\mathbf{q}}|f_\mathbf{q,q}(\mathbf{k})|^2
N_\mathbf{q}(1-N_{\mathbf{q}}),
\end{multline}
where $N_\mathbf{q}=\langle\hat{b}_{\mathbf{q}}^\dagger\hat{b}_{\mathbf{q}}\rangle$ gives the expected population of the lowest band Bloch state with quasimomentum $\mathbf{q}$.
This component is the sum of a contribution due to the average density of the Fermi-Dirac gas and a contribution due to the finite-temperature fluctuations in the average density.
The first-order component, resulting from atoms being excited into unpopulated Bloch states within the lowest band, is 
\begin{equation}
S_{g1} (\mathbf{k})=\sum_{\mathbf{q}\neq\mathbf{p}}|f_\mathbf{q,p}(\mathbf{k})|^2
N_\mathbf{q}(1-N_{\mathbf{p}}).
\label{eq:fermion_first_order}
\end{equation}
The scattering function for scattering involving the higher bands, which we refer to as interband scattering, is given by
\begin{equation}
S_{b} (\mathbf{k}) = N - \sum_{\mathbf{p},\mathbf{q}}
|f_\mathbf{q,p}(\mathbf{k})|^2 N_\mathbf{q}. 
\end{equation}
The above theory extends the treatment in Ref.~\cite{Ruostekoski2009a} through the inclusion of the interband scattering and, further, by consistently working in the Bloch basis we have not made any approximations involving the localization of atoms to lattice sites, allowing our model to be applied over the full range of lattice strengths.

The components of the light scattering are shown in Fig.~\ref{fig:Fermion_sf} as a function of scattering angle for a two dimensional lattice as described in Fig.~\ref{fig:light_scatter}. 
The strongest component is the elastic scattering, which forms sharp diffraction peaks with a diffraction maximum at $\theta=\phi=0$ that scales with $N^2$. The interband component scales with $N$ and is near zero intensity at the diffraction maximum, but can overwhelm the elastic scattering for higher angles and leads to a background level of scattering between the classical diffraction peaks. Neither the elastic term nor the interband term are significantly affected by the temperature or occupation statistics of the gas. The first-order component does change however, for example if the lowest band is completely full with $N=M$, no excitations are available within the lowest band due to Pauli blocking and the first-order component is identically zero, while for $N/M = 0.5$ the contribution from this component scales with $N$. In Fig.~\ref{fig:Fermion_sf}(c) we compare the angular distribution of light scattered for a full lattice with that of a lattice with filling factor $f = N/M = 0.5$. 

The first-order component also decreases as $\theta \rightarrow 0$ due to the Pauli blocking that appears in Eq.~(\ref{eq:fermion_first_order}). Fermions can only scatter into states that are unoccupied and for a particular momentum transfer $\hbar\mathbf{k}$ this restricts the number of atoms that can scatter light. At zero temperature, the only atoms that can be involved in inelastic scattering within the first band are those with momenta $\hbar\mathbf{q}$ such that $\hbar|\mathbf{q}+\mathbf{k}|> p_F$, where $p_F$ is the Fermi momentum. A lower scattering angle corresponds to a lower momentum transfer, reducing the number of atoms that can scatter light and hence the amplitude of light scattering. As temperature increases more final momentum states become available to scatter to throughout the momentum distribution and the low-angle scattering increases with temperature. We further examine the temperature dependence of the light scattering in Sec.~\ref{sec:temp_depend_bog}.


\section{Bosons}

For bosons in an optical lattice at low temperatures $s$-wave interactions between the atoms play an important role. The resulting system dynamics are well described by the Bose-Hubbard model \cite{Jaksch1998b}, which is characterized by the energy of on-site interactions $U$ and the energy associated with tunneling between lattice sites, $J$. For low on-site interaction strength, the atoms form a superfluid. As interaction strength increases a phase transition occurs and the system becomes a Mott insulator. We examine the light scattering from both these phases, beginning with the superfluid regime.

\subsection{Superfluid}
\label{sec:Bogoliubov_theory}

In the superfluid regime the atoms are not localized at lattice sites and, particularly for light scattering, it is most intuitive to expand the atomic field operators in the Bloch basis as we did for the noninteracting fermions. The creation and annihilation operators for the Bloch states now obey the standard boson commutator relations.
The atomic Hamiltonian is then
\begin{equation}
H=\sum_\mathbf{q}E_\mathbf{q}\hat{b}_\mathbf{q}^\dagger\hat{b}_\mathbf{q}+
\frac{1}{2}\sum_{\mathbf{q_1,q_2,q_3,q_4}}U_{\mathbf{q_1},\mathbf{q_2},\mathbf{q_3},\mathbf{q_4}}\hat{b}_\mathbf{q_1}^\dagger\hat{b}_\mathbf{q_2}^\dagger
\hat{b}_\mathbf{q_3}\hat{b}_\mathbf{q_4},
\end{equation}
where
\begin{equation}
U_{\mathbf{q_1},\mathbf{q_2},\mathbf{q_3},\mathbf{q_4}}= \frac{4\pi \hbar^2 a_s M}{m_a} \int d\mathbf{r} \phi^*_\mathbf{q_1}(\mathbf{r})
\phi^*_\mathbf{q_2}(\mathbf{r})\phi_\mathbf{q_3}(\mathbf{r})\phi_\mathbf{q_4}(\mathbf{r}),
\end{equation}
results from the $s$-wave interaction with scattering length $a_s$. Here for notational convenience we have assumed the band index of the Bloch states is contained in the generalized quasimomentum $\mathbf{q}$, and the theory below includes all bands. 

At zero temperature, atoms in a non-interacting Bose gas condense into the zero momentum Bloch state $\phi_\mathbf{0}(\mathbf{r})$. For weak interactions this ground state is perturbed, and the new ground state and elementary excitations of the gas are found using Bogoliubov's theory \cite{Bogolubov1947a}. Extensions of Bogoliubov's theory to Bose gases in optical lattices have been made \cite{vanOosten2001a,Burnett2002a}, but work in the tight binding approximation. Here, by working in the Bloch basis we do not need to make this approximation.

Following Bogoliubov's treatment we replace the operators $\hat{b}_\mathbf{0}$ and $\hat{b}^\dagger_\mathbf{0}$ of the highly populated zero momentum Bloch state by the c-number $\sqrt{N_0}=\sqrt{\langle \hat{b}^\dagger_\mathbf{0}\hat{b}_\mathbf{0}\rangle}$. Then assuming the nonzero-momentum Bloch states each have a population much smaller than $N_0$, we can neglect terms above quadratic order in the Hamiltonian.
The Hamiltonian can then be diagonalized via the Bogoliubov transformation \cite{Bogolubov1947a}
\begin{align}
\hat{b}^\dagger_\mathbf{q} =  u_\mathbf{q}\hat{\beta}^\dagger_\mathbf{q} - v_\mathbf{q}\hat{\beta}_{-\mathbf{q}}, && \hat{b}_\mathbf{q} = u_\mathbf{q}\hat{\beta}_\mathbf{q} - v_\mathbf{q}\hat{\beta}^\dagger_{-\mathbf{q}}, 
\label{eq:bog_trans}
\end{align}
where the Bogoliubov transformation coefficients are given by
\begin{align}
u_\mathbf{q} = \sqrt{\frac{1}{2}\left(\frac{\tilde{E}_\mathbf{q}}{\hbar \omega_\mathbf{q}}+1\right)} && \text{and} &&
v_\mathbf{q} = \sqrt{\frac{1}{2}\left(\frac{\tilde{E}_\mathbf{q}}{\hbar \omega_\mathbf{q}}-1\right)},
\label{Bogoliubov_coef}
\end{align}
with $\tilde{E}_\mathbf{q} = E_\mathbf{q}-E_\mathbf{0}-N_0 U_{\mathbf{0}}+2 N_0 U_{\mathbf{q}}$ and  $U_{\mathbf{q}}=U_{\mathbf{q},\mathbf{0},\mathbf{0},\mathbf{q}}$.
.

The operators $\hat{\beta}_\mathbf{q}^\dagger$ and $\hat{\beta}_\mathbf{q}$ are interpreted as the creation and annihilation operators of noninteracting bosonic quasiparticles with energy $\hbar \omega_\mathbf{q}= \sqrt{\tilde{E}_\mathbf{q}^2-N_0^2 U_{\mathbf{q}}^2}.$
At zero temperature the ground state of the system is the quasiparticle vacuum, while for finite temperature $T$ the population of the quasiparticle modes is given by the Bose-Einstein distribution \cite{Stringari2003, Landau1980}
\begin{equation}
\langle \hat{\beta}^\dagger_\mathbf{q} \hat{\beta}_\mathbf{q'}\rangle = \frac{\delta_{\mathbf{q},\mathbf{q'}}}{e^{\hbar \omega_{\bf{q}}/k_bT}-1}= \delta_{\mathbf{q},\mathbf{q'}}n_\mathbf{q}. 
\end{equation}
Furthermore, we have
$\langle \hat{\beta}_\mathbf{q} \rangle = \langle \hat{\beta}^\dagger_\mathbf{q} \rangle = 0$ and
$\langle \hat{\beta}^\dagger_\mathbf{q} \hat{\beta}_\mathbf{q} \hat{\beta}^\dagger_\mathbf{q'} \hat{\beta}_\mathbf{q'}\rangle =  n_\mathbf{q}n_\mathbf{q'}+\delta_{\mathbf{q},\mathbf{q'}}(n_\mathbf{q}+1)n_\mathbf{q}$,
which we will need to calculate the structure factor.

The above quantities depend implicitly on the condensate number $N_0$, which is in turn restricted by the following relation for the number operator $\hat{N} = \sum_\mathbf{q} \hat{b}^\dagger_\mathbf{q}\hat{b}_\mathbf{q}$, 
\begin{equation}
\langle \hat{N} \rangle = \sum_{\mathbf{q}}N_\mathbf{q}= N_0 + \sum_{\mathbf{q}\neq 0}\left(u_\mathbf{q}^2n_\mathbf{q}+v_\mathbf{q}^2(n_\mathbf{q}+1)\right).
\label{Total_number}
\end{equation}
For a fixed $\langle \hat{N} \rangle = N$, Eqs.~(\ref{Bogoliubov_coef}) and (\ref{Total_number}) must be solved self-consistently. 

As noted above, the Bogoliubov approximation only includes the contributions to the interaction energy proportional to $N_0$, and its validity is therefore limited to the regime where the number of quasiparticles is a small fraction of $N_0$. In this work we limit our calculations so that this fraction is less than one tenth.


\begin{figure}
\centering
\includegraphics{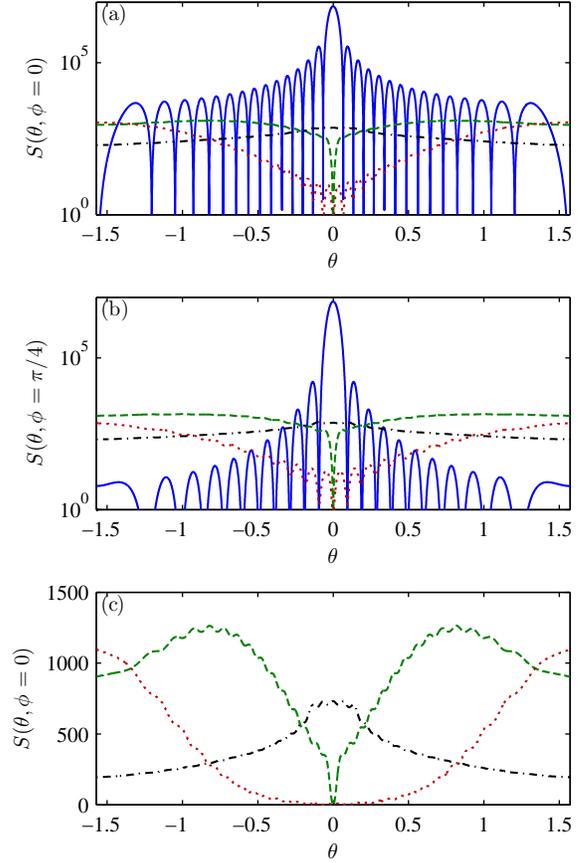}
\caption{(Color online) Comparison of the angular dependence of the components of the static structure factor in the superfluid regime for a $30\times30$ lattice with 2700 atoms. The lattice strengths are $V_x=V_y=3 E_R$ and $V_z= 20 E_R$, leading to parameters $J=0.11 E_R$ and $U=0.17E_R$, the temperature is $0.05 E_R/k_B$ and the condensate population is $N_0=2527$.  (a)  Dependence on $\theta$ of the structure factor components  $S_{g0} (\mathbf{k})$ (solid blue line),  $S_{g1} (\mathbf{k})$ (dashed green line),  $S_{g2} (\mathbf{k})$ (dot-dashed black line) and   $S_{b} (\mathbf{k})$ (dotted red line) for $\phi=0$. (b) As in panel (a) with $\phi=\pi/4$. (c) Inelastic components only.}\label{fig:Comparison_sf_terms}
\end{figure}


As in the fermion case, we assume the temperature is sufficiently low so that all the atoms initially reside in the lowest band of the optical lattice; however atoms may be excited to higher bands by scattering a photon. Examining Eq.~(\ref{Bogoliubov_coef}) we see that if, for the higher bands, $\tilde{E}_\mathbf{q}\gg N_0 U_\mathbf{q}$, then $u_\mathbf{q}\sim 1$ and the quasiparticle excitations are essentially particle-like. This criterion is easily satisfied in the cases we investigate, in which case the only nonzero averages in the static structure factor given by Eq.~(\ref{eq:Static_fact_expand}) are $\langle\hat{b}_{\mathbf{q_1},\mathbf{0}}^\dagger\hat{b}_{\mathbf{q_2},\mathbf{0}}
\hat{b}^\dagger_{\mathbf{q_3},\mathbf{0}}\hat{b}_{\mathbf{q_4},\mathbf{0}}\rangle$ and $\langle\hat{b}_{\mathbf{q_1},\mathbf{0}}^\dagger\hat{b}_{\mathbf{q_2},\mathbf{m}}
\hat{b}^\dagger_{\mathbf{q_2},\mathbf{m}}\hat{b}_{\mathbf{q_3},\mathbf{0}}\rangle = \langle\hat{b}_{\mathbf{q_1},\mathbf{0}}^\dagger\hat{b}_{\mathbf{q_3},\mathbf{0}}\rangle$ for $\mathbf{m}\neq\mathbf{0}$.

We can now evaluate all the expectation values using the Bogoliubov theory for the lowest band, and the quasimomenta $\mathbf{q}$ from this point refer only to the lowest band. We find that light scattering within the lowest band is the sum of three components $S_g (\mathbf{k})=S_{g0} (\mathbf{k})+S_{g1} (\mathbf{k})+S_{g2} (\mathbf{k})$. The zeroth-order component, which does not result in excitations in the lattice, is
\begin{multline}
S_{g0} (\mathbf{k})=\left|\sum_{\mathbf{q}}f_\mathbf{q,q}(\mathbf{k})
N_\mathbf{q} \right|^2 \\+\sum_{\mathbf{q}}|f_\mathbf{q,q}(\mathbf{k})|^2
(u_\mathbf{q}^2+v_\mathbf{q}^2)^2
n_\mathbf{q}(n_{\mathbf{q}}+1).
\end{multline}
As in the fermion case this component results from the average density of the gas and the finite-temperature fluctuations in the average density.
The first-order component, resulting from atoms being excited out of or into the condensate mode, is 
\begin{equation}
S_{g1} (\mathbf{k})=N_0 \sum_{\mathbf{q}\neq\mathbf{0}}|f_\mathbf{0,q}(\mathbf{k})|^2 
\left(u_\mathbf{q}-v_\mathbf{q}\right)^2(2 n_\mathbf{q}+1),
\end{equation}
and the second-order component is
\begin{multline}
S_{g2} (\mathbf{k})=\sum_{\substack{\mathbf{q}\neq\mathbf{0},\mathbf{p}\neq\mathbf{0}\\\mathbf{q}\neq\mathbf{p}}}|f_\mathbf{q,p}(\mathbf{k})|^2
(u_\mathbf{q}u_{\mathbf{p}}+v_\mathbf{q}v_{\mathbf{p}})^2
n_\mathbf{q}(n_{\mathbf{p}}+1)\\
+\frac{1}{2}\sum_{\mathbf{q}\neq\mathbf{0},\mathbf{p}\neq\mathbf{0}}|f_\mathbf{q,p}(\mathbf{k})|^2
(u_\mathbf{q}v_{\mathbf{p}}+v_\mathbf{q}u_{\mathbf{p}})^2
\\\times\left(n_\mathbf{q}n_{\mathbf{p}}+
(n_\mathbf{q}+1)(n_{\mathbf{p}}+1)\right),
\end{multline}
which results from creation and/or destruction involving two quasiparticle modes.
The distribution resulting from interband scattering is given by
\begin{equation}
S_{b} (\mathbf{k}) = N - \sum_{\mathbf{p},\mathbf{q}}
|f_\mathbf{q,p}(\mathbf{k})|^2 N_\mathbf{q}. 
\end{equation}
Again, our theory extends that of previous studies \cite{Rist2010a,Trippenbach2009a} by including the interband scattering and by working consistently in the Bloch basis, with no localization approximations, we have a model that is applicable even for very weak optical lattices.

In Fig.~\ref{fig:Comparison_sf_terms} we compare how the four components of the static structure factor vary as a function of scattering angle.  We see that the classical diffraction pattern is again the dominant feature with peaks of order $N^2$, while the first order and interband components are of order $N$. The second-order component is larger than the first-order and interband components near $\theta=0$, a behavior that persists even at $T=0$.

The behavior of the first- and second-order components reflects the correlations between atoms in the lattice and quantum interference and enhancement play a role in determining their structure. We see in Fig.~\ref{fig:Comparison_sf_terms} that the first-order component vanishes as $\mathbf{k}\rightarrow 0$. This phenomenon has been recorded in Bose liquids \cite{Svensson1980a,Griffin1993a} and in an experiment with a Bose condensed dilute gas, where Bragg spectroscopy was used to probe the structure factor \cite{Stamper-Kurn1999a}. Here the same processes are at work and we can understand the vanishing of the first-order component as interference between two scattering channels. The first channel involves the quantum depletion of the condensate, which consists of pairs of atoms with opposite momentum for finite interparticle interactions \cite{Hua63}. Light scattering can scatter an atom in one of these pairs back into the condensate. This gives the same final state as created by the second  scattering channel where an atom is scattered out of the condensate mode, and the amplitudes of these scattering processes destructively interfere as $\mathbf{k}\rightarrow 0$ (see Ref.~\cite{Stamper-Kurn1999a} for further details).
An interesting feature of the optical lattice case, as opposed to the uniform Bose gas, is that the rate at which the first-order structure factor goes to zero depends on the ratio of $J$ to $U$, which can be tuned by adjusting the height of the lattice or by Feshbach resonance. By adjusting the lattice parameters and examining the light scattering at various angles we can see how the depletion, or alternatively how number squeezing \cite{Burnett2002a}, in the lattice changes. 

Despite involving two quasiparticle modes, the second-order component can still make a significant contribution to the light scattering. This is due to quantum enhancement in the higher order processes as seen in the experiment by Rowen \textit{et al.}~\cite{Rowen2008a}. For this component to be nonzero, noncondensate modes must be populated, either as a result of interparticle interactions or finite temperature.



\subsection{Mott insulator}
\label{sec:Mott_angular}

We now examine light scattering from the Mott insulator. This state is characterized by strong interactions between atoms leading to particle localization and one may expect that the characteristics of light scattered from this state will be significantly different to the superfluid case. 

Again we assume that the temperature is low enough that all the atoms initially reside within the lowest band of the lattice.  Expanding the atomic-field operator in terms of the localized Wannier site basis, $\hat{\Psi}(\mathbf{r}) = \sum_{j,\mathbf{m}}\hat{b}_{j,\mathbf{m}}w_{\mathbf{m}}(\mathbf{r}-\mathbf{r}_j)$, gives the following expression for the static structure factor
\begin{multline}
S(\mathbf{k}) =  \sum_{j_1,j_2,j_3,j_4,\mathbf{m}}f^*_{j_1,\mathbf{0},j_2,\mathbf{m}}( \mathbf{k})f_{j_4,\mathbf{0},j_3,\mathbf{m}}(\mathbf{k})\\\times
\langle\hat{b}^\dagger_{j_1,\mathbf{0}}\hat{b}_{j_2,\mathbf{m}}\hat{b}^\dagger_{j_3,\mathbf{m}}\hat{b}_{j_4,\mathbf{0}}\rangle .
\end{multline} 
where
\begin{equation}
f_{j_1,\mathbf{m},j_2,\mathbf{n}}(\mathbf{k}) = \int d\mathbf{r} w_\mathbf{m}(\mathbf{r}-\mathbf{r}_{j_1})w_\mathbf{n}(\mathbf{r}-\mathbf{r}_{j_2})
e^{i \mathbf{k}\cdot\mathbf{r}}.
\label{eq:wannier_fourier}
\end{equation}
The overlap between lowest-band Wannier functions at different sites is very small in the Mott regime and $f_{j_1,\mathbf{0},j_2,\mathbf{0}}( \mathbf{k})$ can be neglected for $j_1 \neq j_2$.
 
As in the previous cases the terms involving higher bands can be simplified using a sum rule, this time for the Wannier functions. Once again, the description now only involves contributions from the lowest band and we can drop the band index from our notation.
The interband component is then
\begin{equation}
S_b( \mathbf{k}) =  N(1 -|f_{0,0}( \mathbf{k})|^2).
\end{equation}
Here we see that when the site overlap is negligible, the contribution from the higher bands is dependent only on the shape of the site Wannier function and provides no other information about the atoms in the lattice.
The contribution to light scattering from the lowest band is
\begin{equation}
S_g( \mathbf{k}) =  |f_{0,0}( \mathbf{k})|^2\sum_{j,l}
\langle\hat{n}_{j}\hat{n}_{l}\rangle e^{i\mathbf{k}\cdot(\mathbf{r}_j-\mathbf{r}_l)}.
\label{eq:mott_ground_band}
\end{equation}

Deep in the Mott insulator regime the angular dependence of light scattering could be calculated using perturbation theory, 
an approach that has been applied in theoretical work investigating Bragg scattering by Rey \emph{et al.} \cite{Rey2005a} and extended in Rist \emph{et al.} \cite{Rist2010a}.
However, perturbing $J$ from zero leads to corrections to the unperturbed scattering pattern of order $M(J/U)^2$ and corrections to the unperturbed energies of order $J$ \cite{Rey2005a}. When $J \ll U$ these perturbations have negligible effects on the scattering pattern and the scattering pattern is well approximated by that produced when $J=0$. We therefore proceed to calculate the scattering at finite temperature by assuming $J=0$.

To evaluate Eq.~(\ref{eq:mott_ground_band}) in the Mott regime at temperature $T$ we need to calculate $\sum_u \langle\psi_u|\hat{n}_j\hat{n}_l|\psi_u\rangle e^{-E_u/k_B T}/Z$ where $|\psi_u\rangle$ are the eigenstates of the Bose-Hubbard Hamiltonian with energy $E_u$.
Making the approximation $J=0$ \cite{Blakie2007a}, the eigenstates are simply number states $|\{n\}_u\rangle \equiv|\{n^{(u)}_j,j=1,\ldots,M\}\rangle$ with energies $\sum_{j} U n^{(u)}_j (n^{(u)}_j-1)/2$, where the total number of atoms, $\sum_{j} n^{(u)}_j = M n_0$, is fixed with $n_0$ a positive integer. At zero temperature the ground state is simply the number state with $n_0$ atoms at each site.


\begin{figure}
\centering
\includegraphics{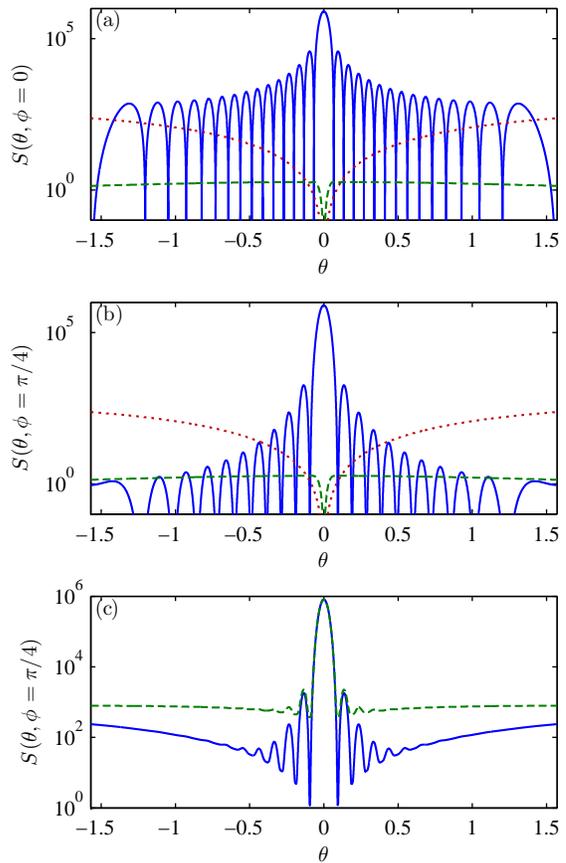}
\caption{(Color online) Angular dependence of components of the static structure factor in the Mott insulator regime for a $30\times30$ lattice with $n_0=1$ and $V_x=V_y=V_z=15 E_R$. The temperature is $0.033E_R/k_B$ ($U = 0.42E_R$) at which point the zero-temperature ground-state proportion is 0.32. (a)  Dependence on $\theta$ of the structure factor components  $S_{g0} (\mathbf{k})$ (solid blue line),  $S_{g1} (\mathbf{k})$ (dashed green line), and  $S_{b} (\mathbf{k})$ (dotted red line) for $\phi=0$. (b) As in panel (a) with $\phi=\pi/4$. (c) Comparison of the Mott insulator structure factor (solid blue line) with that for  the superfluid (dashed green line). For comparison the superfluid state is produced for the same lattice parameters as the Mott insulator, but with the $s$-wave scattering length scaled by a factor of 0.0006, giving a condensate population $N_0 = 807$.}\label{fig:Mott_finite_temp}
\end{figure}


For a translationally invariant system, the energy of these number states remains the same under any permutation of lattice sites.
The eigenstates are then divided into degenerate groups labeled by $v$, where $|\{n\}_v\rangle$ is a representative eigenstate. The other members of the group are got by permutations $P$ of the lattice sites, resulting in $g_v$ different states $|P\{n\}_v\rangle$ each with energy $E_v$. We can use this symmetry to evaluate the required matrix elements of Eq.~(\ref{eq:mott_ground_band}). For each $v$ we have
\begin{equation}
\sum_P\langle P\{n\}_v|\hat{n}_j\hat{n}_l |P\{n\}_v\rangle =\left\{ \begin{aligned} &g_v\langle n^2 \rangle_v &\text{$j=l$ }\\
&\frac{g_v(N^2 - M\langle n^2 \rangle_v)}{M(M-1)} &\text{$j\neq l$,} \end{aligned}\right.
\end{equation}
where $\langle n^2 \rangle_v = \frac{1}{M}\sum_j \left(n^{(v)}_j\right)^2$ is the average over the degenerate group $v$.
This then gives
\begin{multline}
\sum_{j,l}\sum_P\langle P\{n\}_v|\hat{n}_{j}\hat{n}_{l}|P\{n\}_v\rangle e^{i\mathbf{k}\cdot(\mathbf{r}_j-\mathbf{r}_l)}
=\\g_v \left[n_0^2\mathcal{F}(\mathbf{k}) +
\frac{\langle n^2 \rangle_v-n_0^2}{(M-1)}\left(M^2 - \mathcal{F}(\mathbf{k})\right)\right].
\end{multline}
where
\begin{equation}
\mathcal{F}(\mathbf{k}) = \sum_{j,l}e^{i\mathbf{k}\cdot(\mathbf{r}_j-\mathbf{r}_l)} = \prod_{j\in\{x,y,z\}}\frac{\sin^2(M_j  \mathbf{k}_j a/2)}{\sin^2( \mathbf{k}_j a/2)}
\end{equation}
is the classical diffraction pattern from an $M_x\times M_y\times M_z$ array of apertures with inter-site separation $a$ in each dimension.

Finally we find that the lowest-band contribution to the light scatter results in two terms. The first is the classical pattern due to the average density
\begin{equation}
S_{g0}( \mathbf{k}) =  |f_{0,0}( \mathbf{k})|^2 n_0^2 
\mathcal{F}(  \mathbf{k}),
\end{equation}
and the second term is due to the number fluctuations on each site  at finite temperature
\begin{multline}
S_{g1}( \mathbf{k})
=  \frac{|f_{0,0}( \mathbf{k})|^2 \left(M^2-
\mathcal{F}(  \mathbf{k})\right)} {(M-1)Z}\\\times 
\sum_v g_v e^{-E_v/k_B T}(\langle n^2 \rangle_v-n_0^2).
\end{multline}
Neither of these terms result in excitation of the many-body state and the light scattering within the lowest band is purely elastic.

For a number of sites of around 50 this formalism allows us to calculate the light scattering resulting from all possible excitations. We soon see however, that in the temperature range where the zero temperature ground state, $|\{n_j=n_0,j=1,\ldots,M\}\rangle$, is still significantly populated, the only states that play a significant role determining the scattering distribution are the states involving particle-hole excitations. These are the states that have $n_0+1$ atoms at the sites $p_1, p_2, \dots , p_v$ and an equal number of sites $h_1, h_2, \dots, h_v$ with $n_0-1$ atoms. There are $g_v = M!/[(M-2v)!(v!)^2]$ states with $v$ particle-hole pairs and these have energy $E_v = v U$ and site number fluctuations of $(\langle n^2 \rangle_v-n_0^2) = 2v/M$. Restricting our calculations to just these states allows for calculations involving much larger lattices.   

In the limit $T = 0$ the light scattering is purely due to the classical diffraction pattern and interband scatter, no excitation of the lattice occurs within the lowest band, a major difference from the superfluid phase. We can see why this is by considering the simple two site case. Switching from the site basis to the lattice momentum basis using $\hat{b}_\mathbf{q}=\frac{1}{\sqrt{N_s}}\sum_j\hat{b}_j e^{i{\bf q}\cdot {\bf r}_j}$, we see that the Mott state with filling factor one is $\hat{b}_1^\dagger\hat{b}_2^\dagger|0\rangle=1/2((\hat{b}_\mathbf{0}^\dagger)^2-(\hat{b}_\mathbf{q}^\dagger)^2)|0\rangle$, where
$\mathbf{q}=\frac{\pi}{a}(\mathbf{r}_2-\mathbf{r}_1)$. Light scattering involving the wave-vector change $\mathbf{k}=\mathbf{q}$ occurs by two routes,
$\hat{b}_\mathbf{q}^\dagger\hat{b}_\mathbf{0}$ and $\hat{b}_\mathbf{0}^\dagger\hat{b}_\mathbf{q}$. Each of these routes takes the Mott state to the same state, but with opposite signs, and the amplitudes cancel. The lack of excitations hence is a result of interference due to correlations in the lattice momentum distribution.

In Fig.~\ref{fig:Mott_finite_temp} we plot the components of the static structure factor for a Mott insulator at finite temperature. 
The classical diffraction pattern remains the central feature and does not change with temperature. The interband scattering leads to a background level of scattering around the classical diffraction peaks that scales with $N$, which may explain part of the background scattering observed in Ref.~\cite{Weitenberg2011a} that has the same scaling. The interband scattering is dominant over the scattering resulting from the site number fluctuations, which disappear as $T\rightarrow 0$. In Fig.~\ref{fig:Mott_finite_temp}(c) we compare the static structure factor for the Mott insulator with that for the superfluid. The stronger inelastic scattering in the superfluid case leads to higher scattering intensity away from the diffraction maximum, although the difference is not as great as suggested in \L{}akomy \textit{et al.}~\cite{Trippenbach2009a} due to the presence of the interband scattering component.


\section{Temperature dependence of light scattering}
\label{sec:temp_depend_bog}


\begin{figure}
\centering
\includegraphics{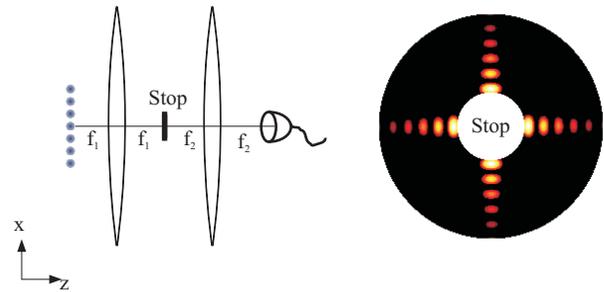}
\caption{(Color online) (Left) Scheme for collecting photons scattered from atoms in an optical lattice to measure the temperature of the gas. A stop placed in the center of the Fourier plane of the first lens blocks the central diffraction peak. A second lens collects the unblocked photons which are then detected. (Right) Outline of the stop on diffraction pattern. }\label{fig:Temp_lens_setup}
\end{figure}



\begin{figure*}
\centering
\includegraphics{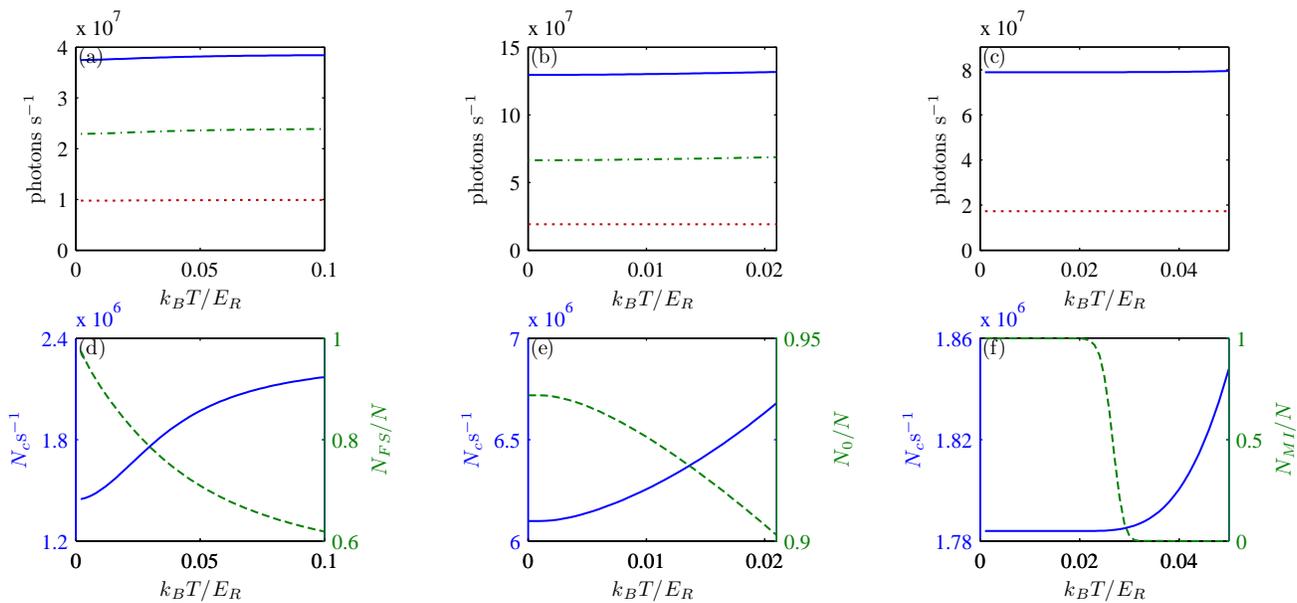}
\caption{(Color online) (a)-(c) Temperature dependence of the total number of photons scattered over all angles (solid line), the number of inelastically scattered photons (dot-dashed line), and the number of photons scattered due to the higher bands only (dotted line) for three many-body states on a $150 \times 150$ lattice. (a) Fermions in a lattice with $V_x=V_y=8 E_R$, $V_z=15 E_R$ and filling factor $f=0.5$. (b)
Superfluid state in a lattice with $V_x=V_y=3 E_R$, $V_z = 20 E_R$ and filling factor $f=1$. (c) Mott insulator state with $V_x=V_y=V_z=20 E_R$ and filling factor $f=1$. Note that for the Mott state the interband scattering makes up all the inelastic scatter.
(d)-(f) Number of photons collected per second by the lens system described in the main text for the same lattices in panels (a)-(c). We also plot on the right axes, with a dashed green line, (d) $N_{FS}/N$, where $N_{FS}$ is the number of atoms with energies below the Fermi energy; (e) $N_0/N$, where $N_0$ is the condensate population for the superfluid; and (f) $N_{MI}/N$, where $N_{MI}$ is the proportion of atoms in the zero temperature ground state for the Mott insulator.
The scattering rates are determined for the D$_2$ line of $^{40}$K in the fermion case and for the D$_2$ line of $^{87}$Rb in the boson cases, each with a detuning of 20$\gamma$ from resonance and a laser intensity of $I_{in}=5Wm^{-2}$.
}\label{fig:All_temp_depend}
\end{figure*}


In all the cases described above components of the static structure factor depend on temperature and we can use this dependence to observe changes in temperature \cite{Trippenbach2009a,Ruostekoski2009a}. In their recent paper, Ruostekoski \textit{et al.}~\cite{Ruostekoski2009a} proposed using non-resonant light scattered from fermions in optical lattices to determine the temperature of the Fermi gas. It is important to test whether their method could also be applied to bosons in optical lattices, as an accurate \textit{in situ} thermometer would be a useful experimental tool \cite{McKay2009a}. Furthermore Ruostekoski \textit{et al.}~neglect excitation of atoms into higher bands in their treatment of light scattering, and the effect of interband scattering on the temperature measurement must be quantified if this method is to be used in experiments.

Following the scheme of Ruostekoski \textit{et al.}, scattered photons can be collected using a system of two lenses as shown in Fig.~\ref{fig:Temp_lens_setup}, where the number of photons detected gives a signal dependent on temperature. As noted by Ruostekoski \textit{et al.}, the elastically scattered photons have little temperature dependence and placing a stop at the center of the Fourier plane of the first lens will improve the signal-to-noise ratio, by blocking the central diffraction peak.


In Figs.~\ref{fig:All_temp_depend}(a)-\ref{fig:All_temp_depend}(c) we plot the temperature dependence of the total number of photons scattered per second over all angles for our three many-body systems. Temperature dependence is weak in all cases. We also plot the proportion of scattering that is inelastic, which is largest for the fermion case and smallest in the Mott insulator. In Figs.~\ref{fig:All_temp_depend}(d)-\ref{fig:All_temp_depend}(f) we plot the temperature dependence of the number of photons collected, $N_c(T)$, by the lens system described above for the three many-body systems. We have assumed that the lens system has a numerical aperture $\sin\theta =0.5$ and the stop blocks light scattered with $\theta< 0.06$. In each case we see an increase in photon number with temperature that could be used to measure the temperature.

As discussed by Ruostekoski \textit{et al.}, inelastic scattering processes heat the sample, and to measure the temperature without perturbing the system significantly we must limit the number of inelastic events $W$ in a single experimental realization. Following Ruostekoski \textit{et al.}~we take $W=0.1N$. To measure the temperature to a useful accuracy it is then necessary to do multiple repetitions $\tau$ of the experiment. The accuracy of the measurement then depends on the Poissonian fluctuations $\sqrt{\tau N_c(T)}$ of the number of photons collected in the $\tau$ repetitions. To achieve an uncertainty $\Delta T$ in the temperature measurement at temperature $T$ we then require $\tau(N_c(T+\Delta T)-N_c(T))\sim\sqrt{\tau N_c(T)}$ \cite{Ruostekoski2009a}. 

For the experimental parameters discussed in Fig.~\ref{fig:All_temp_depend}, determining the temperature with accuracy of $T = 0.05T_F$ ($T_F =0.12 E_R/k_B$) in the fermion case would require 39 repetitions at $T = 0.5T_F$ and 7 repetitions at $T = 0.1T_F $. The presence of the inelastic interband scattering, which is largely temperature independent, reduces the efficiency of the measurement, where for comparison if we had neglected the interband scattering the required number of repetitions would have been 22 and 4. In the superfluid case determining the temperature to accuracy $0.002 E_R/k_B$ would require 23 repetitions at $k_B T =0.02 E_R$ and 197 repetitions at $k_B T =0.004 E_R$. If interband effects were excluded the number of repetitions required would have been 17 and 138, respectively. The temperature independent interband scattering causes the greatest reduction in efficiency in the Mott insulator case, where a temperature measurement at $T = 0.027 E_R/k_B$ would take around 26000 repetitions to achieve an accuracy of 10\%. This results from the weak temperature dependence of the collected photon number relative to the amount of interband inelastic scattering when the zero temperature Mott ground state remains significantly populated, as shown in Fig.~\ref{fig:All_temp_depend}(f). In contrast without the interband contribution all the remaining scattering in our model is elastic and in the limit $U/J \rightarrow \infty$ we would only need one repetition. 

The reduction in efficiency is due mostly to the increased heating resulting from the interband scattering events rather than to a decreased signal-to-noise ratio. This means that filtering out interband photons before collection (as may be possible due to their separation in energy by at least the band gap from the other scattered photons) does not increase the measurement efficiency greatly. In the Mott case above, if we filtered out the interband photons before photon detection the number of repetitions required would be reduced to around 24000. Similarly for the fermion and superfluid cases above the increase in efficiency is 5\% or less.


\section{Conclusions}

We have shown how correlations between atoms in optical lattices are reflected in the way these systems scatter light. For the Fermi-Dirac gas we have seen that Pauli blocking leads to angular dependence of the inelastic light scattering, while for a bosonic superfluid a different angular dependence occurs as a result of interference and enhancement between scattering channels. In the bosonic Mott insulating state inelastic scattering is strongly suppressed by correlations between the atoms in momentum space. We have demonstrated that interband scattering, which has been neglected in other treatments, can significantly change the scattering patterns in a way that is largely independent of correlations within the lattice and temperature. 

We have used our theory to test whether light scattering can be used as a thermometer for ultracold atoms, and have shown that interband scattering reduces the efficiency of temperature measurements based on light scattering, particularly in the Mott insulator case. Our results suggest that using light scattering to measure the temperature of the Mott insulator state will be ineffective, while for the Fermi-Dirac gas and the superfluid this remains a feasible method.

\section*{ACKNOWLEDGMENTS}
We thank Janne Ruostekoski for valuable comments on this work.

%

\end{document}